\documentclass[aps,article,author-year,notitlepage,showpacs]{revtex4-1}
\usepackage{graphicx}
\usepackage{float}
\usepackage[T1]{fontenc}
\usepackage[latin1]{inputenc}
\usepackage{amssymb}
\usepackage{amsmath}
\include{epsf}
\newcommand{\ts}{\vspace*{.2cm}}

\newcommand{\bea}{\begin{eqnarray}}
\newcommand{\eea}{\end{eqnarray}}

\makeatother
\begin{document}

\title{Bethe-Salpeter Equations from the 4PI effective action}

\author{M.E. Carrington$^\dagger$, WeiJie Fu$^\dagger$, T. Fugleberg$^\dagger$, D. Pickering$^*$ and I. Russell$^\dagger$}
\affiliation{$^\dagger$Department of Physics, Brandon University, Brandon, Manitoba, R7A 6A9 Canada and Winnipeg Institute for Theoretical Physics, Winnipeg, Manitoba}

\affiliation{$^*$Department of Mathematics, Brandon University, Brandon, Manitoba, R7A 6A9 Canada}

\date{\today}

\begin{abstract}
In this paper we derive a hierarchy of integral equations from the 4PI effective action which have the form of Bethe-Salpeter equations. We show that, together with the equation of motion for the self-consistent 4-vertex, these integral equations are closed, and that their expansions give infinite series of connected diagrams which have the correct symmetry. 
\end{abstract}

\pacs{11.10.-z, %Field theory
      11.15.Tk, %Other nonperturbative techniques
      11.10.Kk  %Field theories in dimensions other than four
            }

\large
\maketitle

\section{Introduction}
\vspace{5pt}

The $n$-particle irreducible ($n$PI) effective action formalism is a promising approach to study non-perturbative problems  \cite{Jackiw1974,Norton1975}.   
The $n$PI formalism has been used to study finite temperature systems (see for example \cite{Blaizot1999,Berges2005a}),  
non-equilibrium dynamics (see \cite{Berges2001} and references therein), and transport coefficients \cite{Aarts2004,Carrington2008,Carrington2009,Carrington2010b}.  The 4PI theory in particular has been used to obtain the integral equations that correspond to the next-to-leading order contribution to the shear viscosity in scalar $\phi^4$ theory \cite{Carrington2010b}. Gauge invariance of the $n$PI formalism was studied in \cite{Smit2003,Zaraket2004}, and techniques to derive higher order effective actions were developed in \cite{Guo2010,Guo2011}.

To date, very few numerical calculations have been done beyond the level of the 2PI effective theory. Calculations have been done in 3-dimensions for scalar theories using the 4PI effective action \cite{Carrington2012}, and QCD using the 3PI effective action \cite{Moore2012}, but no work has been done in 4-dimensions. One of the problems is that it is not known how to renormalize the $n$PI effective action for $n>2$ in four dimensions. 

The proof of renormalizability of the 2PI effective action in 4-dimensions relies on a non-perturbative 4-vertex which can be obtained from the 2PI effective action using a kind of Bethe-Salpeter (BS) equation \cite{vanHees2002,Blaizot2003,Serreau2005,Serreau2010}.  One re-organizes the equation for the 2-point function by expanding around the known asymptotic solution and writing the divergent piece as an infinite resummation of divergences, and corresponding counter-terms. One can show that the BS equation has exactly the right structure to provide the needed counter-terms. 

Reneormalization is more complicated when self-consistent vertices are involved. 
Higher order BS equations may provide a way to re-organize the integral equations for these vertices so that they can be renormalized. BS type integral equations can be obtained from the Schwinger-Dyson equations \cite{Alkofer2001}. 
In this paper, we look at the BS equations obtained from the 4PI effective action for the symmetric scalar $\phi^4$ theory. Two of these BS equations have been obtained previously, and it has been demonstrated that they have the same structure as the exact renormalization group flow equations, truncated at the level of the second equation \cite{Carrington2013}. In this paper we calculate a complete and closed set of coupled BS equations for the symmetric $\phi^4$ 4PI effective theory, and show that they have the correct symmetry. 

The paper is organized as follows. In section \ref{section:notation} we define our notation. In section \ref{section:nPI} we give a brief review of the $n$PI effective action. In section \ref{section:BS} we review the derivation of the BS equation for the 4-point vertex from the 2PI effective action. 
In section \ref{section:HOBS} we describe the structure of the calculation of higher order BS equations  and in section \ref{section:results} we give the results. 
Our calculations were done using a Mathematica program which can be used generally to manipulate integral equations. The program is available on the web at {\it http://people.brandonu.ca/fuglebergt/files/2012/06/toRun-SH.nb} and is described in section \ref{section:program}. 
In section \ref{section:conclusions} we present our conclusions. Some details are given in the appendix.

\section{Notation}
\label{section:notation}

We work with a scalar field theory with quartic coupling and consider only the symmetric case where the expectation value of the field is zero. 
We define all propagators and vertices with factors of $i$ so that figures look as simple as possible:
lines, and intersections of lines, correspond directly to propagators and vertices, with no additional factors of plus or minus $i$.
The classical action is\footnote{The coupling constant $\lambda$ is imaginary. Using this definition the lines and crosses in Feynman diagrams are propagators and proper vertex functions, as defined in equation (\ref{properDefn}), and diagrams do not carry signs or extra factors of $i$.}
\bea
\label{scl}
S[\varphi]=\int d^dx \bigg(\frac{1}{2}\partial_\mu\varphi(x)\partial^\mu\varphi(x)-\frac{m^2}{2}\varphi(x)^2-\frac{i}{\;4!}\lambda \varphi(x)^4\bigg)\,,
\eea
and the bare propagator is defined:
\bea
\label{bare-prop-def}
G_0^{-1}(x,y) =-i\frac{\delta^2 S[\varphi]}{\delta\varphi(x)\delta\varphi(y)}\,.
\eea
We use a compactified notation in which the space-time coordinates are represented by a single numerical subscript. For example, 
the propagator in equation (\ref{bare-prop-def}) is written $G_{ij}:=G(x_i,x_j)$.  We also use an Einstein convention in which a repeated index implies an integration over space-time variables.
Using this notation we define the generating functionals:
 \bea
 \label{genFcn}
&& Z[J]=\int d\varphi \;{\rm Exp}[i(S+J_i \varphi_i)]\,,\nonumber\\
&& W[J]=-i\,{\rm Ln} Z[J] \,,\nonumber\\
&& \Gamma[\phi]=W[J] - J_i\frac{\delta W}{\delta J_i}\,.
 \eea
The functional $W[J]$ is the generator of connected functions which are defined:
\bea
\label{Wders}
V^c_{123 \cdots k} = \langle\varphi_{x_1}\varphi_{x_2}\varphi_{x_3}\dots\varphi_{x_k}\rangle_c = -(-i)^{k+1}\frac{ \delta^k W}{\delta J_{x_k}\dots \delta J_{x_3}\delta J_{x_2}\delta J_{x_1}}\,.
\eea
The functional $\Gamma[\phi]$ generates 1-line irreducible, or proper, $n$-point functions. Using the notation  $\Gamma = - i\Phi$ they are defined:
\bea
\label{properDefn}
V_{123 \cdots k} =  \frac{\delta^k \Phi[\phi]}{\delta \phi_{x_k}\dots \delta \phi_{x_3}\delta \phi_{x_2}\delta \phi_{x_1}}\,.
\eea

\section{The $n$PI effective action}
\label{section:nPI}

The $n$PI effective action is obtained by taking the $n$th Legendre transform of the generating functional which is constructed by coupling the field to $n$ source terms. Renaming $J$ = ${\cal R}^{(1)}$ we have
\bea
\label{genericGamma}
&& Z[{\cal R}^{(1)},{\cal R}^{(2)},{\cal R}^{(3)},{\cal R}^{(4)},\dots]=\int d\varphi  \;{\rm Exp}[i\,{\cal X}]\,,\\[2mm] &&{\cal X}=S_{cl}[\varphi]+{\cal R}^{(1)}_i\varphi_i+\frac{1}{2}{\cal R}^{(2)}_{ij}\varphi_i\varphi_j + \frac{1}{3!}{\cal R}^{(3)}_{ijk}\varphi_i\varphi_j\varphi_k + \frac{1}{4!} {\cal R}^{(4)}_{ijkl}\varphi_i\varphi_j\varphi_k\varphi_l +\cdots\,,\nonumber\\[4mm]
&&W[{\cal R}^{(1)},{\cal R}^{(2)},{\cal R}^{(3)},{\cal R}^{(4)},\dots]=-i \,{\rm Ln} Z[{\cal R}^{(1)},{\cal R}^{(2)},{\cal R}^{(3)}, {\cal R}^{(4)},\dots]\,,\nonumber\\[4mm]
&&\Gamma[\phi,G,U,V\dots] = W - {\cal R}^{(1)}_i\frac{\delta W}{\delta {\cal R}^{(1)}_i} - {\cal R}^{(2)}_{ij}\frac{\delta W}{\delta {\cal R}^{(2)}_{ij}} - {\cal R}^{(3)}_{ijk}\frac{\delta  W }{\delta {\cal R}^{(3)}_{ijk}} - {\cal R}^{(4)}_{ijkl}\frac{\delta  W }{\delta {\cal R}^{(4)}_{ijkl}}  -\cdots\nonumber
\eea
We consider only the symmetric theory, for which expectation values of an odd number of field operators are zero. 
For future use we give below explicit expressions for the expectation values of products of field operators. Equation (\ref{defcon}) also 
illustrates some of the short-hand notation we will use throughout this paper. Numerical factors in brackets indicate the number of permutations of external legs, only one of which is written explicitly. For example, in the second line of the equation we have written $(3)G_{ij}G_{kl}:= G_{ij}G_{kl} + G_{ik}G_{jl} + G_{il}G_{jk}$. We sometimes suppress space-time indices completely. 
\bea\label{defcon}
%&& 2\frac{\delta  W }{\delta R_{ij}} = 
&&\langle\varphi_i \varphi_j\rangle = G_{ij}+\phi_i \phi_j \,,\\[2mm]
%&& 4!\frac{\delta  W }{\delta R_{ijkl}} = 
&&\langle\varphi_i \varphi_j\varphi_k\varphi_l\rangle = G_{i i^\prime}G_{j j^\prime}G_{k k^\prime}G_{l l^\prime}V_{i^\prime j^\prime k^\prime l^\prime}+(3)G_{ij}G_{jk} \,,\nonumber\\
 \rightarrow &&~  \langle \varphi^4\rangle    = G^4 V+(3)G^2\,,\nonumber\\[2mm]
&&  \langle\varphi_a \varphi_b\varphi_c\varphi_d \varphi_e\varphi_f \rangle = 
(15) G_{a f} G_{b e} G_{c d}
+(15) V_{z_1 z_2 z_3 z_4} G_{a z_1} G_{b z_2} G_{c z_3} G_{d z_4}G_{e f} \nonumber\\
 && ~~~~~~~~~~~~~~~~~~~~~~~ +(10) V_{w_1 w_2 w_3 w_4} V_{z_1 z_2 z_3 z_4} G_{a z_1} G_{b w_1}
   G_{c w_2} G_{d w_3} G_{e z_2} G_{f z_3} G_{w_4 z_4} \nonumber\\
&& ~~~~~~~~~~~~~~~~~~~~~~~ + V^6_{z_1,z_2,z_3,z_4,z_5,z_6} G_{a z_1} G_{b z_2} G_{c z_3}
   G_{d z_4} G_{e z_5} G_{f z_6}\,,   \nonumber\\
\rightarrow &&  ~  \langle \varphi^6\rangle =  (15)G^3 + (15)G^5 V + (10)G^7 V^2 + G^6 V_6\,, \nonumber\\[2mm]
&& \langle \varphi^8\rangle =  (280) G^{10} V^3+ (56) G^9 V V_6 + 315 G^8 V^2 + G^8 V_8 + (28) G^7 V_6 + (210) G^6 V + (105) G^4\,.\nonumber
\eea

\begin{figure}[H]
\begin{center}
\includegraphics[width=11cm]{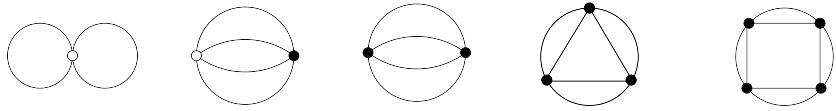}
\end{center}
\caption{The functional $\Phi_2$ for the 5-Loop 4PI effective action. Bare vertices are denoted by open circles and effective vertices are solid dots. \label{fig:PhiAandB-2}}
\end{figure}

The $n$PI effective action is obtained from the last line of Eq. (\ref{genericGamma}). We define $\Phi = i\Gamma$ and $R^{(n)} = i{\cal R}^{(n)}$ and write the result:
\bea
\label{gammaGen}
\Phi[\phi,G,U,V\dots]&& = i W-\frac{1}{2} R_{ij}^{(2)} \langle \varphi_i\varphi_j\rangle - \frac{1}{4!}R_{ijkl}^{(4)} \langle\varphi_i \varphi_j\varphi_k\varphi_l\rangle \dots \\
&&=i S_{cl} - 
    \frac{1}{2} {\rm Tr} \,{\rm Ln}G^{-1}  -
\frac{1}{2} {\rm Tr}\left[ \left(G^0\right)^{-1} G\right] +\Phi_2[G,V\dots]~~+~~{\rm const} \,.\nonumber
\eea
The term $\Phi_2[G,V\dots]$ contains all contributions to the effective action which have two or more loops.
For example, for the 5-Loop 4PI effective action \cite{Carrington2004,Berges2004,Carrington2010b} in the symmetric theory  $\Phi_2$  is shown in figure \ref{fig:PhiAandB-2}.\footnote{All figures are drawn using jaxodraw \cite{jaxo}.}
\par
\begin{figure}[H]
\begin{center}
\includegraphics[width=12cm]{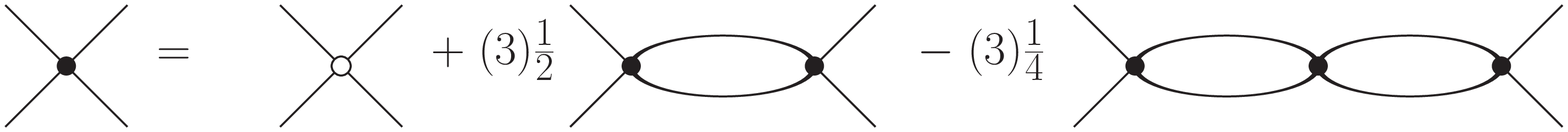}
\end{center}
\caption{Equation of motion for the 4-point function from the 5-Loop 4PI effective action (different permutations of external indices are indicated with a bracketed numerical factor).  \label{fig:Veom}}
\end{figure}

The self consistent propagator and vertex are obtained through the variational principle by solving the equations produced by taking the functional derivative of the effective action and setting the result to zero. For the 4PI effective theory this gives\footnote{The minus sign on the left side of the first equation comes from the fact that the effective action is defined as a functional of the propagator instead of the inverse propagator.}
\bea
\label{Geom}
&& -\frac{\delta \Phi}{\delta G_{xy}}  = - G^{-1}_{xy} + (G^{-1}_0)_{xy}-\Sigma_{xy}=0\,,~~~\Sigma_{xy}=2\frac{\delta\Phi_2}{\delta G_{xy}}\,, \\
\label{Veom}
&& 4! G^{-1}_{xx^\prime} G^{-1}_{xx^\prime} G^{-1}_{xx^\prime} \frac{\delta \Phi}{\delta V_{x^\prime y^\prime w^\prime z^\prime }} = -V_{xywz}+\lambda \delta_{xy}\delta_{xw}\delta_{xz}+4! G^{-1}_{xx^\prime} G^{-1}_{yy^\prime} G^{-1}_{ww^\prime} G^{-1}_{zz^\prime} \frac{\delta \hat\Phi_2}{\delta V_{x^\prime y^\prime w^\prime z^\prime} }   = 0\,.
\eea
The term $- G^{-1}_{xy}$ comes from the 1-loop terms in the effective action and is moved to the other side of the equation to produce the usual form of the Dyson equation. 
%This is shown in figure \ref{fig:SEall-betaS}.  
In equation (\ref{Veom}) the term $-V_{xywz}$ is produced by the basketball (BBALL) diagram in figure \ref{fig:PhiAandB-2} and is moved to the other side to produce the equation of motion shown in figure \ref{fig:Veom}. The second term on the right side comes from the BBALL$_0$ diagram. We use the notation $\hat\Phi_2$ to indicate that the contributions from the BBALL and BBALL$_0$ terms have been removed. 
%\par\begin{figure}[H]
%\begin{center}
%\includegraphics[width=13cm]{SEall-betaS.eps}
%\end{center}
%\caption{Equation of motion for the self-energy from the 4-Loop 4PI effective action. The second line illustrates the notation we use throughout this paper in which different permutations of external indices are indicated with a bracketed numerical factor.  In the third line we have used the equation of motion for the 4-point function which is shown in figure \ref{fig:Veom}. \label{fig:SEall-betaS}}
%\end{figure}
\par

\section{Bethe-Salpeter equation from the 2PI effective action}
\label{section:BS}

It is well known that the 2PI effective action can be used to obtain a 4-point vertex called the Bethe-Salpeter vertex \cite{vanHees2002}. In this section we review the derivation of this equation.  We calculate the functional derivative of the effective action with respect to the 2-point function $G_{kl}$ and the source $R_{ij}$. 
We do the calculation in two different ways and equate the results. First, we use the first line in equation (\ref{gammaGen}) with the expectation values obtained from equation (\ref{defcon}). Differentiating we obtain:
\bea
\label{side1}
\frac{\delta}{\delta R_{ij}} \frac{\delta}{\delta G_{kl}}\Phi = -\frac{1}{4}\big(\delta_{ik}\delta_{jl}+\delta_{il}\delta_{jk}\big)\,.
\eea
Using the second line of equation (\ref{gammaGen}) we obtain:
\bea
\label{side2}
\frac{\delta}{\delta R_{ij}} \frac{\delta}{\delta G_{kl}}\Phi =  \frac{\delta G_{xy}}{\delta R_{ij}}\frac{\delta^2\Phi}{\delta G_{xy}\delta G_{kl}}\,,
\eea
and we write:
\bea
\label{LAM020}
4\frac{\delta^2\Phi}{\delta G_{xy}\delta G_{kl}} && =4\frac{\delta^2(\Phi-\Phi_2)}{\delta G_{xy}\delta G_{kl}}+4\frac{\delta^2\Phi_2}{\delta G_{xy}\delta G_{kl}}\,,\nonumber\\[2mm]
&& = -(G^{-1}_{xk}G^{-1}_{yl}+G^{-1}_{xl}G^{-1}_{yk})+\Lambda_{xykl}\,.
\eea
The inverse propagators in the first term come from the 1-loop terms in the effective action. The second term $\Lambda_{xykl}$ contains all contributions from $\Phi_2$ and has the form of a connected 4-point kernel. 

The last step is to calculate the derivative of the propagator with respect to the source $R$.
We have:
\bea
\label{GderR2}
\frac{\delta G_{xy}}{\delta R_{ij}} &&= \frac{\delta}{\delta R_{ij}} \langle \varphi_x\varphi_y\rangle \,, \nonumber\\
&& = \frac{1}{2}\big(\langle \varphi_x \varphi_y \varphi_i \varphi_j\rangle - \langle \varphi_x\varphi_y\rangle  \langle \varphi_i\varphi_j\rangle \big)\,,  \nonumber\\
&& = \frac{1}{2}\big(G_{ia}G_{jb}G_{xc}G_{yd}M_{abcd}+G_{ix}G_{jy} +G_{iy}G_{jx}\big)\,.
\eea
We will see that the 4-point function $M_{abcd}$ is symmetric with respect to interchange of the first two indices, or the third and fourth indices, or the first set $(a,b)$ and the second set $(c,d)$, but not symmetric with respect to other permutations. For example: $M_{abcd}=M_{bacd}=M_{dcba}\ne M_{acbd}$.  We use the symbol $M$ to denote this vertex, to distinguish it from the self consistent 4-vertex $V$ which is symmetric with respect to all permutations of indices. 

\begin{figure}[H]
\begin{center}
\includegraphics[width=8cm]{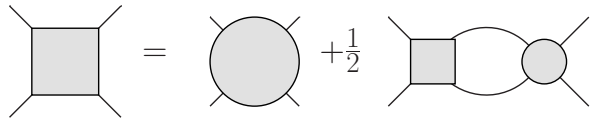}
\end{center}
\caption{Diagrammatic representation of the BS equation in equation (\ref{BSfirst-coord}). Boxes represent the vertex $M$ and circles are the kernel $\Lambda$. \label{fig:standardBS}}
\end{figure}

We substitute equations (\ref{LAM020}) and (\ref{GderR2}) into (\ref{side2}) and set the result equal to the expression obtained in (\ref{side1}). 
In the next section we discuss higher order BS equations which have a much more complicated structure. In order to understand these more complicated calculations, it is useful to look at the substitution described above schematically by suppressing all indices. We obtain:
\bea
-\frac{1}{2}\delta^2&& = \frac{1}{2}(2G^2+G^4 M)\frac{1}{4}(-2 G^{-2}+\Lambda)\,, \\
&& = -\frac{1}{2}\delta^2 -\frac{1}{4} G^2 M +\frac{1}{4}G^2\Lambda+\frac{1}{8}G^2 M G^2 \Lambda \,.\nonumber
\eea
Multiplying on the left by $4G^{-2}$ we obtain the structure of a BS equation:
\bea
M=\Lambda +\frac{1}{2}M G^2 \Lambda \,.
\eea
For the simple example of the 2PI effective action, it is easy to do the calculation (without suppressing indices). The result of performing the substitutions is
\bea
\label{legs}
0=\frac{1}{4} G_{ix} G_{jy}\,\big(-M_{xykl}+\Lambda_{xykl}+\frac{1}{2} M_{xyab} G_{ac}G_{bd} \Lambda_{cdkl} \big)\,.
\eea
Truncating the external legs we obtain the standard form of the BS equation:
\bea
\label{BSfirst-coord}
M_{xykl}=\Lambda_{xykl}+\frac{1}{2} M_{xyab}G_{ac}G_{bd}\Lambda_{cdkl}\,.
\eea
Equation (\ref{BSfirst-coord}) is shown diagrammatically in figure \ref{fig:standardBS}. Throughout this paper we use  circles to denote kernels and boxes are vertices obtained by solving BS integral equations. 
\par
The kernel $\Lambda$ defined in (\ref{LAM020}) contains a series of diagrams which represent $2 \leftrightarrow 2$ amplitudes in $t$- and $u$-channels. The vertex $M$ determined by the BS integral equation gives an infinite resummation of these kernels in the $s$-channel. 

\section{Higher order BS equations}
\label{section:HOBS}

In the previous section we showed how to obtain a 4-point vertex which is symmetric with respect to either pair of indices by taking functional derivatives of the 2PI effective action with respect to $G$ and $R^{(2)}$. In this section we look at functional derivatives of the 4PI effective action with respect to $G$, $V$, $R^{(2)}$ and $R^{(4)}$. 
We will show that these derivatives produce a set of coupled higher order BS equations that is closed (together with the equation of motion for the self-consistent 4-point vertex).

\subsection{Preliminaries}
\label{subsection:prelim}

We list below the derivatives we consider, and our name for the vertex they produce.
\bea
&& \frac{\delta}{\delta R^{(2)}_{ab}}\frac{\delta}{\delta G_{cd}}\Phi ~~ \to ~~ M_{abcd} ~~(M_{22})\,, \\
&& \frac{\delta}{\delta R^{(2)}_{ef}}\frac{\delta}{\delta V_{abcd}}\Phi ~~ \to ~~ M_{ef;abcd}~~(M_{24})\,,\nonumber \\
&& \frac{\delta}{\delta R^{(4)}_{abcd}}\frac{\delta}{\delta G_{ef}}\Phi ~~ \to ~~ M_{abcd;ef}~~(M_{42}) \,,\nonumber\\
&& \frac{\delta}{\delta R^{(4)}_{abcd}}\frac{\delta}{\delta V_{efgh}}\Phi ~~ \to ~~ M_{abcdefgh} ~~ (M_{44})\,. \nonumber
\eea
Each of these vertices is symmetric with respect to the legs that came from the same derivative. 
The vertex $M_{22}$ is the 4PI analogue of the vertex obtained in the previous section (in order to avoid unnecessary indices, we don't introduce additional subscripts to distinguish the 2PI and 4PI effective actions and the propagators and vertices obtained from them). 
When we suppress indices, we use a numerical subscript to distinguish the vertices. The vertices $M_{24}$ and $M_{42}$ have six legs, and the indices that correspond to the symmetric sets of legs are separated by a semi-colon. For $M_{22}$ and $M_{44}$ (the vertex with  eight legs) the semi-colon is not necessary, because it always comes in the middle of the set of indices.  

The equations that produce the BS equations for $M_{22}$, $M_{24}$, $M_{42}$ and $M_{44}$ are:
\bea
\label{M22eqn}
M_{22}\,:~~~\frac{\delta^2\Phi}{\delta R_{ab}\delta G_{cd}} =   \frac{\delta G_{xy}}{\delta R_{ab}}\frac{\delta^2\Phi}{\delta G_{xy}\delta G_{cd}} +  \frac{\delta V_{xywz}}{\delta R_{ab}}\frac{\delta^2\Phi}{\delta V_{xywz}\delta G_{cd}}\,,
\eea

\bea
\label{M24eqn}
M_{24}\,:~~~\frac{\delta^2 \Phi}{\delta R_{ef}\delta V_{abcd}}  = 
\frac{\delta G_{xy}}{\delta R_{ef}}\;\frac{\delta^2\Phi}{\delta G_{xy}\delta V_{abcd}} + \frac{\delta V_{xywz}}{\delta R_{ef}} \; \frac{\delta^2\Phi}{\delta V_{xywz}\delta V_{abcd}}\,,
\eea

\bea
\label{M42eqn}
M_{42}\,:~~~\frac{\delta^2 \Phi}{\delta R_{abcd}\delta G_{ef}}  = 
\frac{\delta G_{xy}}{\delta R_{abcd}}\;\frac{\delta^2\Phi}{\delta G_{xy}\delta G_{ef}} + \frac{\delta V_{xywz}}{\delta R_{abcd}} \; \frac{\delta^2\Phi}{\delta V_{xywz}\delta G_{ef}}\,,
\eea

\bea
\label{M44eqn}
M_{44}\,:~~~\frac{\delta^2 \Phi}{\delta R_{abcd}\delta V_{efgh}}  = 
\frac{\delta G_{xy}}{\delta R_{abcd}}\;\frac{\delta^2\Phi}{\delta G_{xy}\delta V_{efgh}} + \frac{\delta V_{xywz}}{\delta R_{abcd}} \; \frac{\delta^2\Phi}{\delta V_{xywz}\delta V_{efgh}}\,.
\eea

First we calculate the derivatives directly from the inverse transform. These results will be set equal to the result we obtain from the right side of equations (\ref{M22eqn})-(\ref{M44eqn}), and therefore we refer to them as the left sides of each equation. We obtain (see equation (\ref{side1})):
\bea
\label{left-sides}
&& 2\cdot 2\cdot {\rm lhs}^{22}_{abcd} =   -\big(\delta_{ac}\delta_{bd}+\delta_{ad}\delta_{bc}\big)\,,\\
&& {\rm lhs}^{24}_{ef;abcd} =0\,, \nonumber\\
&&2\cdot 4!\cdot {\rm lhs}^{42}_{abcd;ef}  =   - (8) G_{a z_3} G_{b z_2} G_{c z_1} V_{e z_1 z_2 z_3} \delta_{df} - (12) G_{a b} \delta_{cf} \delta_{de} \,,\nonumber\\
&&4! \cdot 4! \cdot {\rm lhs}^{44}_{abcdefgh}  =  - (24) G_{a h} G_{b g} G_{c f} G_{d e}\,.\nonumber
\eea

It is frequently convenient to work with `amputated functional derivatives,' in which the legs left behind by a functional derivative with respect to a 4-vertex or source are truncated. We define an amputated functional derivative:
\bea
\frac{\delta}{\overline{\delta} X_{ijk\dots}} = G^{-1}_{i i^\prime}G^{-1}_{j j^\prime}G^{-1}_{k k^\prime}\dots\frac{\delta}{\delta X_{i^\prime j^\prime k^\prime\dots}}\,.
\eea

\subsection{Kernel notation}
\label{subsection:kernel}

We introduce some notation for the various kernels that we will encounter. 
\bea
\label{LAMdefn}
&& {\bf \Lambda}_{ab cd} = 2\cdot 2\,\frac{\delta}{\delta G_{ab}}\frac{\delta }{\delta G_{cd}} \Phi \,,\\[2mm]
&& {\bf \Lambda}_{ab cd;ef} =2\cdot 4!\, \frac{\delta}{\overline{\delta} V_{abcd}}\frac{\delta}{\delta G_{ef}} \Phi  \,, \nonumber\\[2mm]
&& {\bf \Lambda}_{ef; abcd} = 2\cdot4!\,\frac{\delta}{\delta G_{ef}}\frac{\delta}{\overline{\delta} V_{abcd}} \Phi \,, \nonumber\\[2mm]
&& {\bf \Lambda}_{ab cdefgh} = 4!\cdot4!\,\frac{\delta}{\overline{\delta} V_{abcd}}\frac{\delta}{\overline{\delta} V_{efgh}} \Phi \,.\nonumber
\eea
The derivative $\delta/\overline{\delta} V$ is defined with four inverse propagators that truncate the legs that are left behind by the functional derivative with respect to $V$ so that we obtain an amputated kernel.

A BS equation should contain only connected pieces, and it must have an inhomogeneous term which is just the bare kernel. For the 2PI 4-point function discussed in section \ref{section:BS}, one obtains the form in equation (\ref{BSfirst-coord}) by extracting the disconnected part of the kernel ${\bf \Lambda}_{abcd}$ which is produced by the 1-loop effective action.  Using equation (\ref{LAMdefn}) the kernel in (\ref{LAM020}) is written
\bea
\label{LAM4}
{\bf \Lambda}_{abcd} = -(G^{-1}_{ac}G^{-1}_{bd}+G^{-1}_{ad}G^{-1}_{bc})+\Lambda_{abcd}\,,~~
\eea
where $\Lambda$ comes from the functional derivatives with respect to $G$ acting on the part of the effective action with two and more loops. 
The kernels $\Lambda_{abcd;ef}$, $\Lambda_{ef;abcd}$ and $\Lambda_{abcdefgh}$ do not receive contributions from the 1-loop part of the effective action, and the structure of the corresponding BS equations is different. This is discussed in detail in the following sections. For the moment, we make the definitions:
\bea
\label{LL}
 {\bf \Lambda}_{abcd;ef}&& = \Lambda_{abcd;ef}\,,~~ {\bf \Lambda}_{ef;abcd} = \Lambda_{ef;abcd}\,,\\
 {\bf \Lambda}_{abcdefgh}&& = 4!\cdot 4!\,\bigg(\frac{\delta^2 \Phi_{\rm bball}}{\overline{\delta} V_{abcd}\overline{\delta} V_{efgh}}
 +\frac{\delta^2 (\Phi-\Phi_{\rm bball})}{\overline{\delta} V_{abcd} \overline{\delta} V_{efgh}}\bigg)  =  - (24)G^{-1}_{ae}G^{-1}_{bf}G^{-1}_{cg}G^{-1}_{dh} + \Lambda_{abcdefgh}\,.\nonumber
\eea
When suppressing indices we use the notation: $\Lambda_{abcd}=\Lambda_{22}$, $\Lambda_{ef;abcd}=\Lambda_{24}$, $\Lambda_{abcd;ef}=\Lambda_{42}$ and  $\Lambda_{abcdefgh}=\Lambda_{44}$.

We give below the results for the kernels obtained from the 5-Loop 4PI effective action. 

The kernel $\Lambda_{abcd}$ is:
\bea
\label{LAM2c2}
&&\Lambda_{abcd}=
V^0_{abcd}+
(4)\frac{1}{2}G_{z_2 z_3} G_{z_1 z_4} V^0_{a d z_2 z_4} V_{b c z_1 z_3}-(2)\frac{1}{2}G_{z_2 z_3} G_{z_1 z_4} V_{a d z_2 z_4} V_{b c z_1 z_3}\\
%2-loop
~&&+(4)\frac{1}{2}G_{z_3 z_4} G_{z_1 z_5} G_{z_2 z_6} G_{z_7 z_8} V_{a z_4 z_6 z_8} V_{c z_1 z_2 z_3} V_{b d z_5 z_7} 
+ (2)\frac{1}{4} G_{z_4 z_5} G_{z_1 z_6} G_{z_2 z_7} G_{z_3 z_8} V_{z_1 z_2 z_4 z_8} V_{a d z_5 z_7} V_{b c z_3 z_6}\nonumber\\
%3-loop
~&&-(2)\frac{1}{4} G_{z_3 z_5} G_{z_2 z_7} G_{z_4 z_8} G_{z_9 z_{10}} G_{z_6 z_{11}} G_{z_1 z_{12}} V_{a z_5 z_8 z_{12}} V_{b z_3
   z_7 z_{11}} V_{c z_2 z_6 z_9} V_{d z_1 z_4 z_{10}} \nonumber \\
~&&-(4)\frac{1}{4} G_{z_3 z_5} G_{z_2 z_6} G_{z_4 z_8} G_{z_9 z_{10}} G_{z_7 z_{11}} G_{z_1 z_{12}} V_{z_1 z_4 z_7 z_{10}} V_{a z_5
   z_8 z_{12}} V_{c z_2 z_9 z_{11}} V_{b d z_3 z_6} \nonumber \\
~&&-(2)\frac{1}{8} G_{z_2 z_4} G_{z_3 z_6} G_{z_7 z_9} G_{z_8 z_{10}} G_{z_5 z_{11}} G_{z_1 z_{12}} V_{z_1 z_3 z_5 z_9} V_{z_4 z_6
   z_{10} z_{12}} V_{a d z_2 z_8} V_{b c z_7 z_{11}} \,.\nonumber
\eea
The first line is from the EIGHT+BBALL$_0$+BBALL graphs, the second line is from LOOPY, and the third, fourth and fifth lines are from LOOPY5.

It is not obvious that the two 6-point kernels are symmetric partners of each other. In fact, the kernel obtained by differentiating with respect to $G$ first and $V$ second contains additional terms relative to the one obtained by taking the derivatives in the opposite order, but one can show that all of these terms cancel using the equation of motion for the self-consistent vertex. 
%As is the case for the 2PI effective action, the kernels contain only $t$- and $u$-channels. 
The result is:
\bea
\label{LAM4c2}
\Lambda_{abcd;ef}
&& = (8)G^{-1}_{ae}V^0_{bcdf}-(8)G^{-1}_{ae}V_{bcdf} \\
&& +(24)\frac{1}{2}G^{-1}_{ae}V_{bfz_1z_2}G_{z_1z_3}G_{z_2z_4}V_{z_3z_4cd}+(6)V_{abez_1}G_{z_1z_2}V_{z_2cdf}\nonumber\\
&& -(24)\frac{1}{4}G^{-1}_{ae}V_{bfz_1z_2}G_{z_1z_3}G_{z_2z_4}V_{z_3z_4z_5z_6} G_{z_5 z_5^\prime}G_{z_6 z_6^\prime} V_{z_5z_6cd}\nonumber\\
&& -(12)\frac{1}{2}V_{abz_1z_2}G_{z_1z_3}G_{z_2z_4}V_{z_3z_4z_5e}G_{z_5z_6}V_{cdz_6f}\,,\nonumber
\eea
where the terms in the first, second and third lines are, respectively, from the (BBALL$_0$+BBALL), LOOPY,  LOOPY5 diagrams in the effective action. It is easy to see that the disconnected pieces (the terms that contain inverse propagators) cancel using the equation of motion for the self-consistent vertex (see figure \ref{fig:Veom}). The kernel $\Lambda_{ef;abcd}$ has no contributions from the BBALL$_0$ and BBALL diagrams, and no disconnected pieces from the LOOPY and LOOPY5 diagrams, and the diagrams that are obtained are the symmetric partners of the surviving terms in equation \ref{LAM4c2}.
 
The kernel $\Lambda_{abcdefgh}$ contains disconnected terms. We give the result below and discuss later the role of the disconnected terms. 
\bea
\label{LAM4c4}
\Lambda_{abcdefgh} && = 
(72) V_{a b e f} G^{-1}_{ch} G^{-1}_{dg}  -\frac{1}{2} (72) G_{z_2 z_3} G_{z_1 z_4} V_{a b z_3 z_4} V_{e f z_1 z_2}
   G^{-1}_{ch} G^{-1}_{dg} - (18) V_{a d g h} V_{b c e f}\,.
\eea
The first term comes from the LOOPY diagram and the second and third are both from LOOPY5 (the contribution from BBALL was already extracted in equation (\ref{LL})). 
%All of these terms are disconnected (although they do not necessarily produce disconnected contributions to the BS equation). 

\begin{figure}[H]
\begin{center}
\includegraphics[width=12cm]{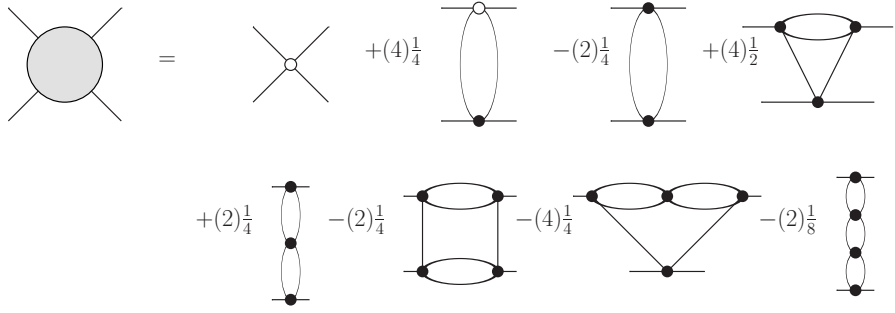}
\end{center}
\caption{The 4-point kernel from the 5-loop 4PI effective action. \label{fig:LAM4}}
\end{figure}

\begin{figure}[H]
\begin{center}
\includegraphics[width=8cm]{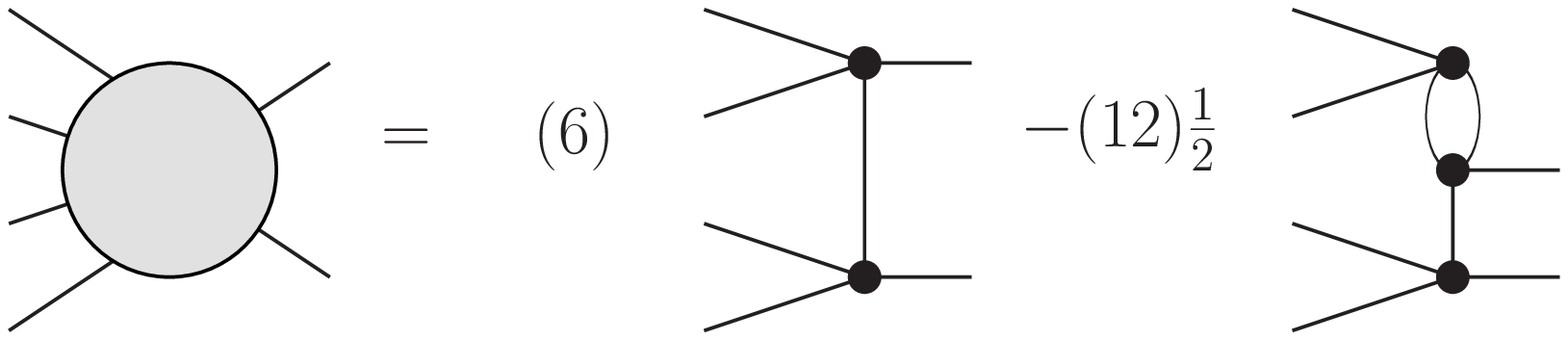}
\end{center}
\caption{The 6-point kernel from the 5-loop 4PI effective action. \label{fig:LAM6}}
\end{figure}

\begin{figure}[H]
\begin{center}
\includegraphics[width=13cm]{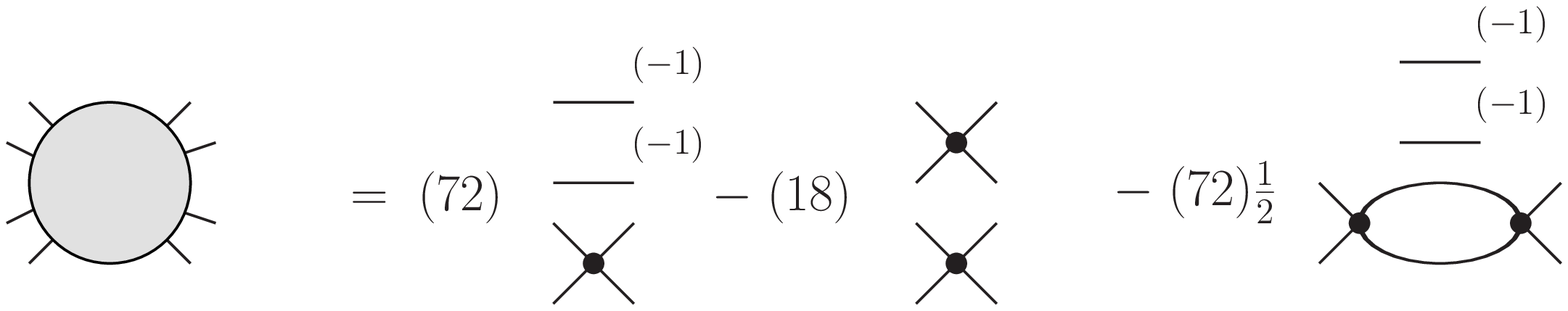}
\end{center}
\caption{The 8-point kernel from the 5-loop 4PI effective action. The lines with superscript (-1) represent inverse propagators that connect indices on either side of the vertex. \label{fig:LAM8}}
\end{figure}

Equations (\ref{LAM2c2}), (\ref{LAM4c2}) and (\ref{LAM4c4}) are shown in figures \ref{fig:LAM4}, \ref{fig:LAM6} and \ref{fig:LAM8}. 
\par
\par
\par
%From these figures it is clear that the kernels introduced in this section (like the 2PI 4-point kernel) contain only $t$- and $u$- channels. The BS equations will resumm these kernels in the $s$-channel. 

\subsection{Source Derivatives}
\label{subsection:source-derivatives}

\begin{figure}[H]
\begin{center}
\includegraphics[width=11cm]{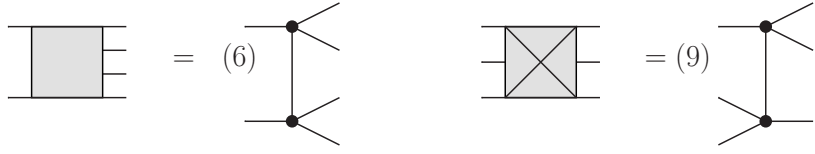}
\end{center}
\caption{The 6-point vertices $\tilde M_{24}$ and $\hat V$ (represented by the box with the cross in it).  \label{fig:M6tilde}}
\end{figure}

In this section we calculate the four amputated derivatives $\delta G_{xy}/\overline{\delta} R_{ab}$, $\delta V_{xywz}/\overline{\delta} R_{ab}$, $\delta G_{xy}/\overline{\delta} R_{abcd}$ and $\delta V_{xywz}/\overline{\delta} R_{abcd}$.  The calculation is tedious but straightforward. 
The derivative $\frac{\delta G_{xy}}{\delta R_{ab}}$ was calculated in section \ref{section:BS} and the corresponding amputated derivative is easily obtained from equation (\ref{GderR2}) as
\bea
\label{chopGdR2}
2\frac{\delta G_{xy}}{\overline{\delta} R_{ab}} = \delta_{xa}\delta_{yb} + \delta_{xb}\delta_{yc}+G_{x i}G_{yj}M_{ijab}\,.
\eea
The method to calculate 
$\frac{\delta G_{xy}}{\overline{\delta} R_{abcd}}$ is exactly analogous. The result contains three kinds of terms, two of which produce disconnected and 1PR contributions to the BS equation, and one which does not.  We group these three kinds of terms separately, and in the next section we will show that the first two cancel in BS equations and only the term marked `surv' survives. 
\bea
\label{divideGdR4}
&& 4! \frac{\delta G_{xy}}{\overline{\delta} R_{abcd}}  = (GdR_4)_{\rm surv}+(GdR_4)_{\rm 1PR}+(GdR_4)_{\rm disco}\,, \\
&& (GdR_4)_{\rm surv} = \tilde M_{abcd;z_1z_2}G_{z_1x}G_{z_2y}\,,\nonumber\\
&& (GdR_4)_{\rm 1PR} = (4)M_{a z_1z_2z_3}G_{z_3z_4}V_{z_4bcd}G_{z_1x}G_{z_2y}+(8)\delta_{ay}G_{xz_1}V_{z_1bcd} = 2(4)\frac{\delta G_{xy}}{\overline{\delta}R_{az_1}}G_{z_1z_2}V_{z_2bcd} \,,\nonumber\\
&& G^4(GdR_4)_{\rm disco} = 2(6)G_{cd}\frac{\delta G_{xy}}{\delta R_{ab}} \,. \nonumber
\eea
The result on the right side of the last line is for the non-amputated derivative, which is indicated by the factor $G^4$ on the left side. The vertex $\tilde M_{abcd;z_1z_2}$ contains the $t$- and $u$-channel tree graphs and is defined as:
\bea
\label{eqn:M6tilde}
\tilde M_{abcd;ef}&& =M_{abcd;ef}+ G_{z_3 z_4} V_{a b e z_4} V_{c d f z_3}+G_{z_3 z_4} V_{a d e z_4} V_{b c f z_3}+G_{z_3 z_4} 
V_{a cf z_3} V_{b d e z_4}+G_{z_3 z_4} V_{ab f z_4} V_{c d e z_3} \nonumber\\
&& +G_{z_3 z_4} V_{a d f z_4} V_{b c e z_3}+G_{z_3 z_4} V_{a c e z_3} V_{b d f z_4}\,,\nonumber  \\
&& = M_{a b c d; e f} +(6) G_{z_1 z_2} V_{a d f z_2} V_{b c e z_1} \,.\nonumber 
\eea
It is symmetric with respect to the first four indices and the second two indices, and is shown in figure \ref{fig:M6tilde}. 

The derivatives of the proper vertex functions are slightly harder to calculate. We give some details for the calculation of $\delta V/\overline{\delta} R^{(2)}$ in Appendix \ref{appendix:VR}. The result is:
\bea 
\label{chopVderR2}
2\frac{\delta V_{xyzw}}{\overline{\delta} R_{ab}} && =  M_{ab;xyzw} + (6) G_{z_3 z_4}  V_{a xy  z_3} V_{b zw z_4} = \tilde M_{ab;xyzw}\,,
\eea
where $\tilde M_{ab;xyzw}$ is the symmetric partner of $\tilde M_{xyzw;ab}$.

\begin{figure}[H]
\begin{center}
\includegraphics[width=12cm]{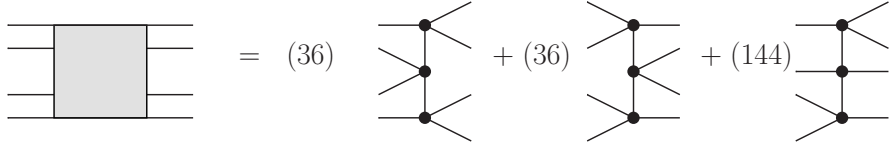}
\end{center}
\caption{The 8-point vertex $\tilde M_{44}$. \label{fig:M8tilde}}
\end{figure}

The calculation of $\delta V/\overline{\delta} R^{(4)}$ is the most difficult and we give only the result. Again, we divide it into three pieces (the result on the right side of the last line is for the non-amputated derivative as indicated by the factor $G^4$ on the left side). The result is:
\bea
\label{divideVdR4}
&& 4! \frac{\delta V_{xyzw}}{\overline{\delta} R_{abcd}}  = (VdR_4)_{\rm surv}+(VdR_4)_{\rm 1PR}+(VdR_4)_{\rm disco}\,,  \\
&& (VdR_4)_{\rm surv} = \tilde M_{abcdxyzw}+(24)G^{-1}_{ax}G^{-1}_{by}G^{-1}_{cz}G^{-1}_{dw} + (72)G^{-1}_{ax}G^{-1}_{by}V_{cdzw}+(18)V_{abxy}V_{cdzw}+(16)G^{-1}_{ax}\hat V_{bcd;yzw} \,,\nonumber\\
&& (VdR_4)_{\rm 1PR} = (4)\tilde M_{xyzw;dz_2} G_{z_2z_1}V_{z_1abc} = 2(4)\frac{\delta V_{xyzw}}{\overline{\delta}R_{dz_2}} G_{z_2z_1}V_{z_1abc} \,,\nonumber\\
&& G^4(GdR_4)_{\rm disco} = 2(6)G_{cd}\frac{\delta V_{xyzw}}{\delta R_{ab}} \,,\nonumber
\eea
where we have defined:
\bea
\label{eqn:M8tilde}
\tilde M_{abcdefgh} &&=    
M_{abcdefgh} + (144) G_{z_1 z_2} G_{z_3 z_4} V_{a h z_2 z_3} V_{b f g z_4} V_{c d e z_1} + (36) G_{z_1 z_2}
   G_{z_3 z_4} V_{e f z_2 z_3} V_{a d h z_4} V_{b c g z_1} \\
 &&  + (36) G_{z_1 z_2} G_{z_3 z_4} V_{c d z_2 z_3} V_{a g h z_4} V_{b e f z_1}\,,\nonumber\\
\hat V_{abc;def} &&=   G_{z_3 z_4} V_{a b d z_4} V_{c e f z_3}+G_{z_3 z_4} V_{a b e z_4} V_{c d f z_3}+G_{z_3 z_4} V_{a d e z_4} V_{b c f z_3}+G_{z_3 z_4} V_{a
   c f z_3} V_{b d e z_4} \nonumber\\
   && +G_{z_3 z_4} V_{a b f z_4} V_{c d e z_3}+G_{z_3 z_4} V_{a d f z_4} V_{b c e z_3}+G_{z_3 z_4} V_{a c e z_3} V_{b
   d f z_4}+G_{z_3 z_4} V_{a e f z_4} V_{b c d z_3}+G_{z_3 z_4} V_{a c d z_3} V_{b e f z_4} \nonumber\\
    && = (6) G_{z_1 z_2} V_{a d f z_2} V_{b c e z_1}+ (3) G_{z_1 z_2} V_{a e f z_2} V_{b c d z_1}\,.\nonumber
   \eea
The vertex $\tilde M_{abcdefgh}$ is symmetric with respect to the first four indices, and the second four indices, and interchange of the first four and the second four. The vertex $\hat V_{abcefg}$ is symmetric with respect to the first three indices, and the second three indices, and interchange of the first three and the second three. 
The definitions in equations (\ref{eqn:M6tilde}) and (\ref{eqn:M8tilde}) are shown in figures \ref{fig:M6tilde} and \ref{fig:M8tilde}.
\par
\par

\section{Results}
\label{section:results}

We obtain the BS equations by substituting the definitions of the kernels, and the results for the derivatives of the field variables with respect to the sources, into the master equations ((\ref{M22eqn}) - (\ref{M44eqn})). 
%The first two equations were derived previously in Ref. \cite{Carrington2013} using slightly different notation. 
When performing these substitutions, one cannot of course ignore different permutations of indices, but must include explicitly all terms. This makes the calculation prohibitively difficult to do by hand and we therefore use a Mathematica program, which is discussed in section \ref{section:program}. 

The boxes in this section represent the vertices $M_{22}$, $\tilde M_{24}$, $\tilde M_{42}$ and $\tilde M_{44}$, and the circles denote $\Lambda_{22}$, $\Lambda_{24}$, $\Lambda_{42}$ and $\Lambda_{44}$. 

\subsection{$M_{22}$ equation}
\label{subsection:M22}
\begin{figure}[H]
\begin{center}
\includegraphics[width=11cm]{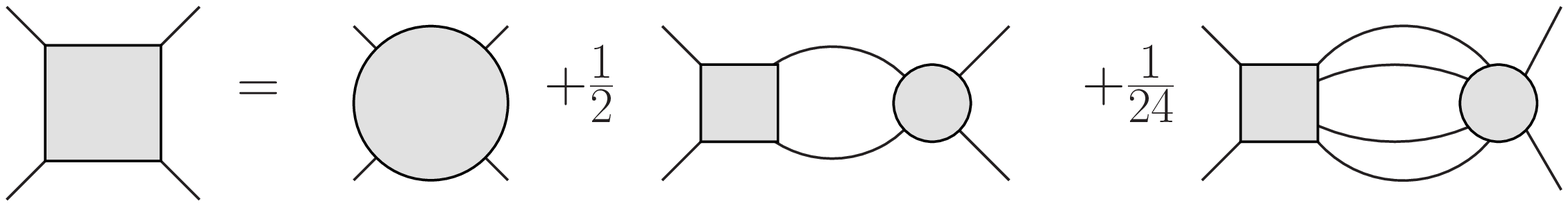}
\end{center}
\caption{The BS equation for the vertex $M_{22}$. \label{fig:M22}}
\end{figure}

Comparing equations (\ref{side2}) and (\ref{M22eqn}), it it clear that the BS equation for the 4PI vertex $M_{22}$ differs from the corresponding 2PI one by the addition of one extra term (the last term on the right side of (\ref{M22eqn})). From (\ref{legs}), the BS equation for $M_{22}$ is obtained from the master equation in (\ref{M22eqn}) by multiplying by 4 and using an amputated $R^{(2)}$ derivative instead of the non-amputated one. The new term in (\ref{BSfirst-coord}) is thus:
\bea
4 \frac{\delta V_{xyzw}}{\overline \delta R_{ab}} G_{x x^\prime}G_{y y^\prime}G_{z z^\prime}G_{w w^\prime} \frac{\delta^2\Phi}{\overline{\delta} V_{x^\prime y^\prime w^\prime z^\prime}\delta G_{cd}} = \frac{1}{24}\tilde M_{ab;xyzw}G_{x x^\prime}G_{y y^\prime}G_{z z^\prime}G_{w w^\prime} \Lambda_{x^\prime y^\prime z^\prime w^\prime;cd}\,,
\eea
where we have used (\ref{LAMdefn}), (\ref{LL}) and (\ref{chopVderR2}) in the last line. Note that the amputated derivative with respect to the internal vertex introduces four internal propagators: $\delta/\delta V_{xyzw} = G_{x x^\prime}G_{y y^\prime}G_{z z^\prime}G_{w w^\prime} \delta/\overline{\delta} V_{x^\prime y^\prime w^\prime z^\prime}$. The full BS equation for the 4PI $M_{22}$ vertex is represented in figure \ref{fig:M22}. The vertex $M_{22}$ is coupled to the vertex $M_{24}$ which is calculated in the next section. 
\par

\subsection{$M_{24}$ equation}
\label{subsection:M24}
\begin{figure}[H]
\begin{center}
\includegraphics[width=11cm]{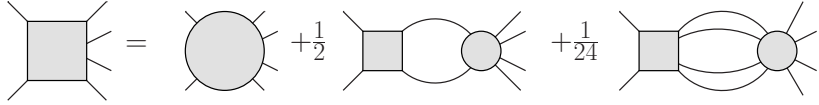}
\end{center}
\caption{The BS equation for the vertex $M_{24}$. \label{fig:M24}}
\end{figure}

We start by rewriting the master equation (\ref{M24eqn}) in the form:
\bea
0  = 
\frac{\delta G_{xy}}{\overline{\delta} R_{ab}}\,(2\cdot 4!)\frac{\delta^2\Phi}{\delta G_{xy}\overline{\delta} V_{cdef}} +\frac{1}{12} \frac{\delta V_{xywz}}{\overline{\delta} R_{ab}} \; G_{xx^\prime}G_{yy^\prime}G_{zz^\prime}G_{ww^\prime} \,(4!\cdot 4!) \frac{\delta^2\Phi}{\overline{\delta} V_{x^\prime y^\prime w^\prime z^\prime}\overline{\delta} V_{cdef}}\,.
\eea
The zero on the left side of this equation comes from (\ref{left-sides}). The amputated derivatives with respect to functions with external indices ($R_{ab}$ and $V_{cdef}$) are obtained by multiplying on each side by the appropriate number of inverse propagators, and the amputated derivative with respect to the internal vertex brings with it the four internal propagators. We have also multiplied by an overall factor $(2\cdot 4!) = \frac{1}{12}(4!\cdot 4!)$ for future convenience. 
We substitute the components of this equation from the previous two sections using equations (\ref{LAMdefn}), (\ref{LL}), (\ref{chopGdR2}) and (\ref{chopVderR2}). 
We give the result schematically, suppressing all indices:
\bea
0 &&= \frac{1}{2}(2\delta^2+G^2 M_{22})\Lambda_{24}+\frac{1}{12}\frac{1}{2}\tilde M_{24}G^4 (-(24)G^4+\Lambda_{44})\,,  \\
&& = \Lambda_{24}+\frac{1}{2}M_{22}G^2\Lambda_{24} - \tilde M_{24} + \frac{1}{24} \tilde M_{24} G^4 \Lambda_{44} \,,\nonumber\\
\to~~ &&\tilde M_{24} = \Lambda_{24}+\frac{1}{2}M_{22}G^2\Lambda_{24}  + \frac{1}{24} \tilde M_{24} G^4 \Lambda_{44} \,.\nonumber
\eea
Notice that although the inhomogeneous term comes from the first functional derivative in the master equation, as was the case for the $M_{22}$ equation, the left side of the BS equation now comes from the second term. The equation has formally the standard BS form, and is shown in figure \ref{fig:M24}.  Since the kernel $\Lambda_{44}$ contains disconnected pieces  the form shown in figure \ref{fig:M24} is somewhat miss-leading, but one can show that even though the kernel $\Lambda_{44}$ is disconnected, the BS vertex $M_{24}$ is not. This is discussed in more detail in section \ref{subsection:expansions}. 

\par

\subsection{$M_{42}$ equation}
\label{subsection:M42}
The master equation for the vertex $M_{42}$ is equation (\ref{M42eqn}). 
The left side is given in (\ref{left-sides}) and can be written more conveniently as:
\bea
\label{lhs42}
&& 24\,{\rm lhs}^{42}_{abcd;ef} = 24\,{\rm lhs}^{42A}_{abcd;ef} + 24\,{\rm lhs}^{42B}_{abcd;ef} \,,\\
&& 24\,{\rm lhs}^{42A}_{abcd;ef} = 2(6) G_{cd}\,{\rm lhs}^{22}_{abef} \,,\nonumber\\
&& 24\,G^{-1}_{aa^\prime}G^{-1}_{bb^\prime}G^{-1}_{cc^\prime}G^{-1}_{dd^\prime} {\rm lhs}^{42B}_{a^\prime b^\prime c^\prime d^\prime ;ef} = 2(4) G^{-1}_{a u}G^{-1}_{zw} {\rm lhs}^{22}_{efuw}G_{zt}V_{tbcd}\,. \nonumber
\eea
We also use that the derivatives $\delta G/\delta R^{(4)}$ and $\delta V/\delta R^{(4)}$ can be divided into three pieces using equations (\ref{divideGdR4}) and (\ref{divideVdR4}). 

We consider first the part of the left side marked ${\rm lhs}^{42A}$ in (\ref{lhs42}) and the  pieces of each derivative marked `disco' in (\ref{divideGdR4}) and (\ref{divideVdR4}). Subsituting into the master equation and including an overall factor (4!) we obtain from these terms:
\bea
\label{M42eqn-b}
2(6) G_{cd}\,{\rm lhs}^{22}_{abef}   = 
2(6)G_{cd}\frac{\delta G_{xy}}{\delta R_{ab}} 
\;\frac{\delta^2\Phi}{\delta G_{xy}\delta G_{ef}} + 2(6)G_{cd}\frac{\delta V_{xyzw}}{\delta R_{ab}}  \; \frac{\delta^2\Phi}{\delta V_{xywz}\delta G_{ef}}\,.
\eea
This result is just $2(6)G_{cd}$ times the BS equation for the 4-vertex $M_{abef}$, and is therefore identically zero. 

Next we look at the part of the left side marked ${\rm lhs}^{42B}$ in (\ref{lhs42}) and the  pieces of the derivatives marked `1PR' in (\ref{divideGdR4}) and (\ref{divideVdR4}). Subsituting into the amputated form of the master equation and including an overall factor (4!) we obtain:
\bea
&& 2(4) G^{-1}_{a w_2}G^{-1}_{z_1w_2} {\rm lhs}^{22}_{efw_1w_2}G_{z_1z_2}V_{z_2bcd} \\
&&= 2(4)\frac{\delta G_{xy}}{\overline{\delta}R_{az_1}}G_{z_1z_2}V_{z_2bcd} \frac{\delta^2\Phi}{\delta G_{xy}\delta G_{ef}}  
+ 2(4)\frac{\delta V_{xyzw}}{\overline{\delta}R_{dz_1}} G_{z_1z_2}V_{z_2abc} \frac{\delta^2\Phi}{\delta V_{xywz}\delta G_{ef}}\,,\nonumber\\
\to ~&& 2(4)\bigg[-G^{-1}_{a w_2}G^{-1}_{z_1w_2} {\rm lhs}^{22}_{efw_1w_2} + \frac{\delta G_{xy}}{\overline{\delta}R_{az_1}} \frac{\delta^2\Phi}{\delta G_{xy}\delta G_{ef}}  
+ \frac{\delta V_{xyzw}}{\overline{\delta}R_{dz_1}}  \frac{\delta^2\Phi}{\delta V_{xywz}\delta G_{ef}}\bigg]G_{z_1z_2}V_{z_2bcd}=0\,.\nonumber
\eea
The quantity in square brackets is the amputated master equation that produces the 4-vertex $M_{az_1ef}$ and is therefore identically zero. 

\begin{figure}[H]
\begin{center}
\includegraphics[width=13cm]{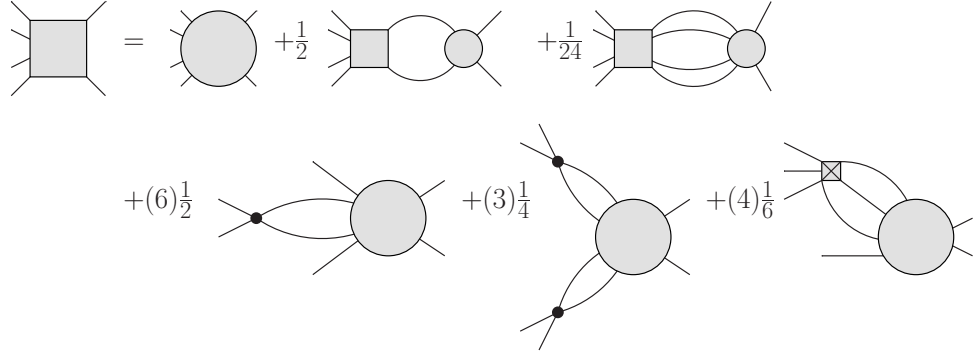}
\end{center}
\caption{The BS equation for the vertex $M_{42}$. \label{fig:M42}}
\end{figure}

Finally we collect the surviving terms and show they have the form of a BS equation. Collecting the terms marked `surv' in (\ref{divideGdR4}) and (\ref{divideVdR4}) and including an additional factor of 2 we have:
\bea 
&& 0= 2 \tilde M_{abcdz_1z_2}G_{z_1x}G_{z_2y} \frac{\delta^2\Phi}{\delta G_{xy}\delta G_{ef}} \\
&&  + 
\big[\tilde M_{abcdxyzw}+(24)G^{-1}_{ax}G^{-1}_{by}G^{-1}_{cz}G^{-1}_{dw} + (72)G^{-1}_{ax}G^{-1}_{by}V_{cdzw}+(18)V_{abxy}V_{cdzw}+(16)G^{-1}_{ax}\hat V_{bcd;yzw}\big]\frac{\delta^2\Phi}{\delta V_{xywz}\delta G_{ef}}\,.\nonumber
\eea
We give below the form of the equation with indices suppressed, in order to see the structure. 
Using equations (\ref{LAMdefn}), (\ref{LAM4}) and (\ref{LL}) we have
\bea
&& \frac{\delta^2\Phi}{\delta G_{xy}\delta G_{ef}} \to \frac{1}{4}(-(2)G^{-2}+\Lambda_{22})\,,\\
&& \frac{\delta^2\Phi}{\delta V_{xywz}\delta G_{ef}} \to \frac{1}{2\cdot 4!}G^4\Lambda_{42}\,,\nonumber
\eea
and substituting gives:
\bea
0= 2 \tilde M_{42} G^2 \frac{1}{4}(-(2)G^{-2}+\Lambda_{22})+2\big[\tilde M_{44}+(24)G^{-4} + (72)G^{-2}V+(18)V^2+(16)G^{-1}\hat V\big]\frac{1}{2\cdot 4!}G^4\Lambda_{42}\,.\nonumber\\
\eea
Rearranging we can write
\bea
\tilde M_{42}= \Lambda_{42}+2 \tilde M_{42} G^2 \frac{1}{4} \Lambda_{22} + 2\big[\tilde M_{44} + (72)G^{-2}V+(18)V^2+(16)G^{-1}\hat V\big]\frac{1}{2\cdot 4!}G^4\Lambda_{42}\,.
\eea
Thus we find that for the $M_{42}$ equation the left side of the BS equation comes from the first functional derivative in the master equation, and the inhomogeneous term comes from the second functional derivative. The result is shown in figure \ref{fig:M42}.
\par

\subsection{$M_{44}$ equation}
\label{subsection:M44}

\begin{figure}[H]
\begin{center}
\includegraphics[width=15cm]{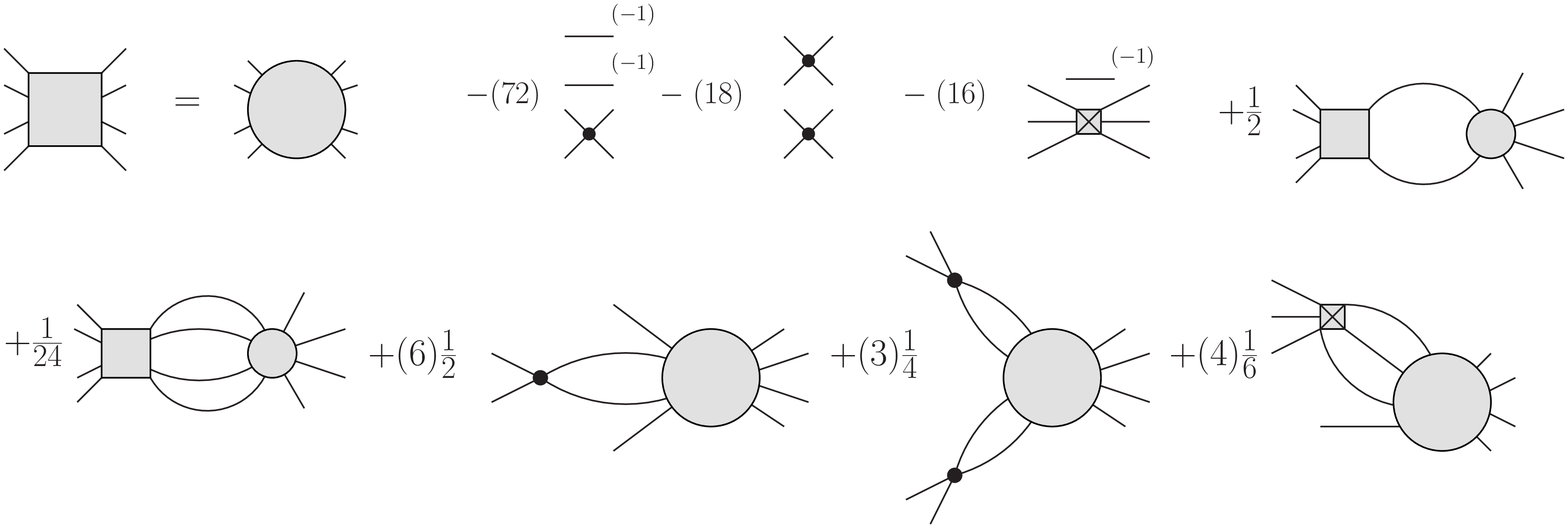}
\end{center}
\caption{The BS equation for the vertex $M_{44}$. The lines with superscript (-1) represent inverse propagators that connect indices on either side of the vertex. \label{fig:M44}}
\end{figure}

Once again, we start from the appropriate master equation (\ref{M44eqn}) and divide the source derivatives into three pieces. In this case it is easy to see that the terms marked `disco' and `1PR' cancel, because the internal BS equation that is produced is the $M_{24}$ equation for which the left side is zero (see equation (\ref{left-sides})). Substituting the kernels from equation (\ref{LL}), the surviving terms have the form:
\bea
&& 4! G^{-1}_{a a^\prime}G^{-1}_{bb^\prime}G^{-1}_{cc^\prime}G^{-1}_{dd^\prime} {\rm lhs}^{44}_{a^\prime b^\prime c^\prime d^\prime efgh}  = (GdR_4)^{\rm surv}_{abcdxy}\frac{1}{2\cdot 4!}\Lambda_{xy;efgh} \\
~~&& +(VdR_4)^{\rm surv}_{abcdxyzw} G_{xx^\prime} G_{yy^\prime} G_{zz^\prime} G_{ww^\prime}\frac{1}{4!\cdot 4!}\big(-(24)G^{-1}_{x^\prime e}G^{-1}_{y^\prime f}G^{-1}_{z^\prime g}G^{-1}_{w^\prime h}+\Lambda_{x^\prime y^\prime z^\prime w^\prime efgh}\big)\,, \nonumber
\eea
where the terms in the round bracket are produced when two functional derivatives with respect to $V$ act on the effective action, the first one coming from the BBALL diagram (see equation (\ref{LL})).
Substituting the results for $(VdR_4)_{\rm surv}$ and $(VdR_4)_{\rm surv}$ from equations (\ref{divideGdR4}) and (\ref{divideVdR4}), and suppressing indices and multiplying by an additional factor (24) we obtain:
\bea
-4!\cdot 4! \cdot \frac{1}{4!}G^{-4} = \frac{1}{2}\tilde M_{42}G^2 \Lambda_{24} + 4!\big[\tilde M_{44}+(24)G^{-4} + \cdots \big]G^4 \frac{1}{4!4!}(-24 G^{-4}+\Lambda_{44})\,.
\eea
The terms in the square bracket come from $(VdR_4)_{\rm surv}$. The first term in the square bracket, together with the first term in the round bracket, provides the left side of the BS equation for the vertex $M_{44}$. 
The second term in the square bracket, with the first term in the round bracket, cancels the left side of the master equation. The second term in the square bracket and the second term in the round bracket give the bare kernel which provides the inhomogeneous term in the BS equation. Including all terms and indices we obtain:
\bea
\tilde M_{44}&& = \Lambda_{44} -(72)G^{-1}_{ae} G^{-1}_{bf} V_{cdgh}  
-  (18) V_{abef}V_{cdgh} - (16) G^{-1}_{ae} \hat V_{bcd;fgh} + \frac{1}{2}\tilde M_{abcd;z_1z_2}G_{z_1z_3}G_{z_2z_4}\Lambda_{z_3z_4;cdef}\\
&& +\frac{1}{4!}\tilde M_{abcd;z_1z_2z_3z_4}G_{z_1w_1}G_{z_2w_2}G_{z_3w_3}G_{z_4w_4}\Lambda_{w_1w_2w_3w_4efgh} + \frac{(6)}{2} V_{abz_1w_1}G_{z_1z_2}G_{w_1w_2}\Lambda_{cd z_2w_2 efgh} \nonumber\\
&& + \frac{(3)}{4}V_{abz_1z_2}V_{cdz_3z_4}G_{z_1w_1}G_{z_2w_2}G_{z_3w_3}G_{z_4w_4}\Lambda_{w_1w_2w_3w_4efgh}+\frac{(4)}{6}\hat V_{abc;z_1z_2z_3} G_{z_1w_1}G_{z_2w_2}G_{z_3w_3} \Lambda_{d w_1w_2w_3efgh}\,.\nonumber
\eea
This result is shown in figure \ref{fig:M44}. 
\par

\subsection{Expansions}
\label{subsection:expansions}

The $M_{44}$ equation derived in the previous section does not have the standard form of a BS equation because it contains disconnected pieces. This is to be expected, since the kernel $\Lambda_{44}$ also contains disconnected pieces at all orders (see figure \ref{fig:LAM8}). Notice however that the equations for the vertices $M_{24}$ and $M_{42}$ are connected, because the disconnected terms that enter through $\Lambda_{44}$ and $M_{44}$ are only horizontally disconnected, and therefore do not generate disconnected terms in the BS equations for $M_{24}$ and $M_{42}$. 

We note also that the results for the vertices $M_{24}$ and $M_{42}$ in figures \ref{fig:M24} and \ref{fig:M42} do not appear to be symmetric. As a further check of our equations, we have expanded the $M_{24}$ and $M_{42}$ equations to two loop order and checked that the vertices are symmetric partners of each other: $M_{abcd;ef}=M_{ef;abcd}$. 

The expanded $M_{44}$ equation satisfies $M_{abcdefgh}=M_{efghabcd}$ but it does contain a disconnected piece, although this might cancel at the level of the 6PI effective action. The expansion of these integral equations is done using a Mathematica program which is described in the next section. 

\section{Program}
\label{section:program}

The process of expanding integral equations is straightforward, but when dealing with a set of coupled equations it becomes prohibitively tedious  beyond 1-loop (or at most 2-loops). In this section, we describe how it can be done using Mathematica. 

The basic problem is that one must change the names of the dummy variables when performing a substitution. Consider for example a simple integral equation containing one $s$-channel graph:
\bea
\label{ex1}
\text{V}_{abcd}=\text{V}^0_{abcd}+
\frac{1}{2} \text{V}^0_{ab\text{zum1}\text{zum2}}G_{\text{zum1}\text{zum3}} G_{\text{zum2}\text{zum4}}
 V_{\text{zum3}\text{zum4}cd}\,.
\eea
%In the last term, the variables after the semi-colon are the dummies. 
In order to substitute the equation for $V$ into itself, one must change the names of the dummy variables. 
To perform a second substitution, one must change them again. 
%This process quickly generates a huge number of terms. 
In all but the most trivial integral equations, many terms are generated which are copies of other diagrams with differently named dummy variables. We have written a program to perform the substitutions and combine terms that correspond to copies of the same diagram. This program is available on-line at {\it http://people.brandonu.ca/fuglebergt/files/2012/06/toRun-SH.nb}. We describe below how the program works. Further help can be obtained from any of the authors by email. 

The first section contains a list of allowed vertices. Additional vertices can be added by the user. 
For example, if the following vectors are defined:
\bea
&& {\rm namee2}=\{{\rm del},G\}\,, \nonumber\\
&& {\rm namee3}=\{U^0,U\}\,, \nonumber\\
&& {\rm namee4}=\{V^0,V\}\,, \nonumber
\eea
one can construct Feynman diagrams (like the ones represented in equation (\ref{ex1})) using propagators $G$, 3-vertices $U_0$ and $U$, and 4-vertices $V_0$ and $V$. One can define vertices with 3, 4, 5, 6 and 8 legs. The notation del is used for the delta function, and must always be included in the list namee2.

There are two sets of dummy indices that can be used. The canonical set is:
\bea
\{{\rm zum1},{\rm zum2},{\rm zum3}\cdots {\rm zum69}\}\,,
\eea
and the second set is 
\bea
\{{\rm z1},{\rm z2},{\rm z3}\cdots {\rm z69}\}\,.
\eea 
Allowed external variables are $\{a,b,c,d,\cdots m\}$. 

When performing the first substitution, one uses a substitution rule in which the dummy variables are shifted from the canonical set to the second set. For example, the first substitution of equation (\ref{ex1}) into itself produces:
\bea
\label{ex2}
{\rm V}_{abcd}[{\rm step2}]\;&&\; = V^0_{abcd}+\frac{1}{2}
   G_{\text{z1}\text{z3}} G_{\text{z2}\text{z4}}
   V^0_{ab\text{z1}\text{z2}}
   V^0_{\text{z3}\text{z4}cd}\\
\;&&\;+\frac{1}{4} G_{\text{z1}\text{z3}} G_{\text{z2}\text{z4}}
   G_{\text{zum1}\text{zum3}} G_{\text{zum2}\text{zum4}}
   V^0_{ab\text{z1}\text{z2}} V_{\text{zum3}\text{zum4}cd}
   V^0_{\text{z3}\text{z4}\text{zum1}\text{zum2}}\,.\nonumber
\eea
One can check that all dummy variables appear twice using the module tallyCheck. For example, tallyCheck[${\rm V}_{abcd}[{\rm step2}]$] produces:
\bea
&& {\rm external ~variables~ are},~ \{{\rm a,b,c,d}\}\,,\nonumber\\
&& {\rm dummy ~variables~ are}, ~\{{\rm z1,z2,z3,z4,zum1,zum2,zum3,zum4}\}\,.\nonumber
\eea
\noindent For expressions involving large numbers of terms, tallyCheck checks 500 randomly selected terms, and warns the user that only 500 terms have been checked. 

\ts

The module shifterM shifts all dummy variables to members of the canonical set. For example, shifterM[${\rm V}_{abcd}[{\rm step2}]$] produces:
\bea
\label{ex3}
{\rm V}_{abcd}[{\rm step3}] &&=V^0_{abcd}+\frac{1}{2}
   G_{\text{zum1}\text{zum3}} G_{\text{zum2}\text{zum4}}
   V^0_{ab\text{zum1}\text{zum2}}
   V^0_{cd\text{zum3}\text{zum4}}\\
&& +\frac{1}{4} G_{\text{zum1}\text{zum3}} G_{\text{zum2}\text{zum4}}
   G_{\text{zum5}\text{zum7}} G_{\text{zum6}\text{zum8}}
   V^0_{ab\text{zum1}\text{zum2}} V_{cd\text{zum7}\text{zum8}}
   V^0_{\text{zum3}\text{zum4}\text{zum5}\text{zum6}}\,.\nonumber
\eea
Further substitutions can be performed in an interative manner. 

For the simple example given above, this procedure is sufficient to obtain the correct topologies and the correct symmetry factors. For more complicated integral equations however, another problem arises. 
In order to understand the difficulty, consider the equation shown in figure \ref{fig:LAM4} which contains four diagrams with vertical bubbles (the 2nd, 3rd, 5th and 8th diagrams on the right side of the figure). If this equation is expanded using the equation of motion for the vertex $V$ (figure \ref{fig:Veom}), a diagram with two vertical bubbles is produced in both the first and second iterations. These terms should be combined to obtain the correct symmetry factor, but the expansion process can produce different dummy variables assignments. 
One must identify terms in the expansion which correspond to the same topology, which involves looking at different permutations of the dummy variables that appear in the final expression. Beyond the 1- or 2-loop level this becomes impractical because the factorials that result from looking at permutations are so large. Our method to handle this problem is described below.  

\begin{figure}[H]
\begin{center}
\includegraphics[width=3cm]{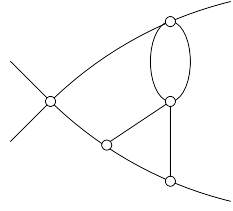}
\end{center}
\caption{An example of a diagram which can be produced with different dummy variable assignments. \label{fig:example3}}
\end{figure}

Consider the diagram in figure \ref{fig:example3}. 
\par
\noindent Iterations can produce copies of this diagram of the form:
\bea
{\rm orig} && = G_{\text{zum1}\text{zum12}} G_{\text{zum10}\text{zum6}}
   G_{\text{zum11}\text{zum4}} G_{\text{zum13}\text{zum3}}
   G_{\text{zum14}\text{zum9}} G_{\text{zum2}\text{zum8}}
   G_{\text{zum5}\text{zum7}} \\
&& ~ \times\,U_{d\text{zum13}\text{zum14}}
   U_{\text{zum1}\text{zum2}\text{zum3}} V_{ab\text{zum11}\text{zum12}}
   V_{c\text{zum4}\text{zum5}\text{zum6}}
   V_{\text{zum10}\text{zum7}\text{zum8}\text{zum9}} \,,\nonumber\\
{\rm copy} && = G_{\text{zum1}\text{zum6}} G_{\text{zum10}\text{zum12}}
   G_{\text{zum11}\text{zum4}} G_{\text{zum13}\text{zum3}}
   G_{\text{zum14}\text{zum9}} G_{\text{zum2}\text{zum8}}
   G_{\text{zum5}\text{zum7}}\nonumber \\
&& ~ \times\, U_{d\text{zum13}\text{zum14}}
   U_{\text{zum10}\text{zum2}\text{zum3}} V_{ab\text{zum11}\text{zum12}}
   V_{c\text{zum4}\text{zum5}\text{zum6}}
   V_{\text{zum1}\text{zum7}\text{zum8}\text{zum9}} \,.\nonumber
\eea
The copy is the same as the original if one interchanges the dummy variables zum1 and zum10. We need to get Mathematica to identify copies. We refer to this process as `tracking.'

The first step in tracking is to reduce the set of indices corresponding to each vertex. For example, for 3-point vertices one uses the rule $U_{\text{zum1zum2zum3}} \to U_{\text{zum1}}{\rm del}_{\text{zum3zum1}}{\rm del}_{\text{zum2zum1}}$. We refer to this procedure as `unUing.'  Although unUing reduces the number of dummy indices and makes the permutation factors manageable, it also destroys all information about the order of the original set of indices. For vertices which are symmetric with respect to all permutations of their indices this is not a problem, but for BS vertices one has only symmetry with respect to permutations of indices on either side of the semi-colon. At the end of this section we explain how to deal with this problem. 
After unUing, the second step is to consider all permutations of the remaining dummy variables and combine terms which are copies of each other. 
Finally, in order to perform the next iteration, one must recover the form in which all vertex variables are specified. This can be done by keeping track of which terms in the original expression survive in the final one. 
%if one inputs: unU0U=copy1+copy2
%the output is:
%\bea
%unUed = V(a) V(c) U(d) U(\text{zum2}) V(\text{zum1}) \text{del}(ab) G_{ac}
%   G_{a\text{zum2}} G_{c\text{zum1}}^2 G_{d\text{zum1}} G_{d\text{zum2}}
%   G_{\text{zum1}\text{zum2}}+V(a) V(c) U(d) U(\text{zum1}) V(\text{zum2})
%   \text{del}(ab) G_{ac} G_{a\text{zum1}} G_{c\text{zum2}}^2
%   G_{d\text{zum1}} G_{d\text{zum2}} G_{\text{zum1}\text{zum2}}
%\eea
%In some cases, two copies will be combined when the unUed form is produced, but in this case, as seen above, the two terms contain different dummy variables assignments. 
For the example above, if one inputs: toTrack=orig + copy the output is:
\bea
{\rm TRACKED} &&= 2 G_{\text{zum1}\text{zum11}} G_{\text{zum10}\text{zum12}}
   G_{\text{zum13}\text{zum7}} G_{\text{zum14}\text{zum6}}
   G_{\text{zum2}\text{zum4}} G_{\text{zum3}\text{zum9}}
   G_{\text{zum5}\text{zum8}} \nonumber\\
&&~\times\,  U_{d \text{zum5}\text{zum6}}  U_{\text{zum2}\text{zum7}\text{zum8}} V_{ab\text{zum3}\text{zum4}}
   V_{c\text{zum10}\text{zum11}\text{zum9}}
   V_{\text{zum1}\text{zum12}\text{zum13}\text{zum14}}\,,
\eea
which shows that the two equivalent terms have been combined with a coefficient 2. 

\ts

There are two additional sections to the program which are useful. 

When expanding integral equations, one wants to truncate at some loop order.  The module loopOrderFunction separates a list of terms by the number of loops. 
The output is loop0, loop1, $\cdots$ loop11, and an error is printed if diagrams with more than 11 loops are present. 

It is frequently useful to look at expressions in which terms that correspond to different permutations of external indices are combined. We refer to this process as `dumbing.' The structure of this problem is the same as when looking for equivalent permutations of internal indices. This section of the program does not perform unUing automatically, but it can be done separately if the number of internal indices is large. In practice, unUing is usually necessary (depending on the number of terms in the input expression) if the number of dummy variables is greater than four. We look at the simple example of two 1-loop diagrams, one in the $s$-channel and one in the $t$-channel. 
The input is:
\bea
\label{dumbperm}
{\rm dumb}~~&& =\frac{1}{2}
 V_{ab\text{zum1}\text{zum2}}  G_{\text{zum1}\text{zum3}}  G_{\text{zum2}\text{zum4}}
    V_{\text{zum3}\text{zum4}cd}  \\
&& + \frac{1}{2} V_{ad\text{zum1}\text{zum2}} G_{\text{zum1}\text{zum3}} G_{\text{zum2}\text{zum4}}
  V_{\text{zum3}\text{zum4}cb} \,.\nonumber
\eea
The second term in this equation can be obtained from the first by exchanging $b \leftrightarrow d$. 
The output of the program is:
\bea
{\rm DUMB} = 
\frac{1}{2} \text{am}(2)V_{ad\text{zum1}\text{zum2}} G_{\text{zum1}\text{zum3}} G_{\text{zum2}\text{zum4}}
    V_{bc\text{zum3}\text{zum4}}\,.
\eea
The numerical factor that corresponds to the number of permutations of external legs that have been combined is preserved in the factor am(2). 

\ts

Now we discuss the issue of index ordering. 
The procedures we have described above will not give the correct result unless it is true that all vertices are symmetric under all permutations of external legs, and for Bethe-Salpeter vertices, this is not true. 

First we consider external indices. The program that combines permutations of external legs takes an input variable `symtype.' If this variable is not defined, all possible permutations of the external variables are considered. If the input is, for example, symtype=abXcd, dumb = $\cdots$ then only permutations of the variables $\{ab\}$ and $\{cd\}$ are considered (but not permutations like $a\leftrightarrow c$ or $b\leftrightarrow d$). In the example discussed above, if we use symtype=abXcd the two terms in equation (\ref{dumbperm}) will not be combined, since they are $b\leftrightarrow d$ permutations of each other.

A long list of choices for symtype is possible (for example abXcdef, abcdXefgh, etc). In each case, the program combines permutations of legs from indices on each side of the X, but not indices on either side of the X. It is simple for the user to add additional choices for symtype if desired. 

The problem of ordering is more serious when looking at permutations of internal indices. As mentioned above, the unUing procedure is necessary to reduce the number of internal indices, but it destroys the information about the order of the original set of indices, and different orderings of internal indices are indistinguishable to the program. In the calculations done in this paper, we have overcome this problem by assigning variables in alphabetic order in such a way that the correct index order can be recovered. For example, the 6-vertex $\tilde M_{{\rm zum1}{\rm zum2}{\rm zum3}{\rm zum4};ef}$ will emerge in the result `TRACKED' as $\tilde M_{ef{\rm zum1}{\rm zum2}{\rm zum3}{\rm zum4}}$, but based on the original variable assignment it is easy to see how to restore the proper order to the indices. This trick will work for any vertices in which the two sets of indices on either side of the semi-colon are one set which contain only external variables and one set which contains only dummy variables, which is always the case for BS equations.

\section{Conclusions}
\label{section:conclusions}

We have written a Mathematica program which is useful generally for expanding and manipulating integral equations. For example, a previous version of our program was used in Refs. \cite{Guo2010,Guo2011} to compare the structure of the Schwinger-Dyson equations and the $n$PI equations of motion. 
In this paper we have used the program to derive a coupled set of higher order BS equations for the symmetric scalar $\phi^4$ theory at the level of the 4PI effective action. We have shown that this set of equations, together with the equation of motion for the self-consistent vertex, is closed, and that they have the right symmetry. The renormalizability of the 2PI effective theory depends on the use of the 2PI BS equation, and we therefore expect that our 4PI Bethe-Salpeter equations will play an important role in the proof of renormalizability of the 4PI effective theory. This work is in progress. 

\section*{Acknowledgements}

This work was supported by the Natural and Sciences and Engineering Research Council of Canada.

\appendix

\section{Derivation of the functional derivative $\delta V/\delta R$}
\label{appendix:VR}

In this appendix we give some details of the calculation of the derivative $\delta V_{xywz}/\delta R_{ab}$. We start from 
\bea
\label{VderR2-a}
\frac{\delta}{\delta R_{ab}} V_{xywz}=\frac{\delta}{\delta R_{ab}}\bigg(G^{-1}_{xx^\prime}G^{-1}_{xx^\prime}G^{-1}_{xx^\prime}G^{-1}_{xx^\prime}V^c_{x^\prime y^\prime w^\prime z^\prime}\bigg)\,.
\eea
and separate the contributions from the derivative acting on the inverse propagators and the connected vertex function. If the derivative acts only on the vertex function we have:
\bea
\label{VderR2part1}
~~&&G^{-1}_{xx^\prime}G^{-1}_{yy^\prime}G^{-1}_{ww^\prime}G^{-1}_{zz^\prime}\frac{\delta}{\delta R_{ab}} V^c_{x^\prime y^\prime w^\prime z^\prime} \nonumber\\
&& = G^{-1}_{xx^\prime}G^{-1}_{yy^\prime}G^{-1}_{ww^\prime}G^{-1}_{zz^\prime} \frac{\delta}{\delta R_{ab}}\big( \langle \varphi_{x^\prime} \varphi_{y^\prime} \varphi_{w^\prime} \varphi_{z^\prime}\rangle - \langle\varphi_{x^\prime}\varphi_{y^\prime} \rangle \langle\varphi_{w^\prime}\varphi_{z^\prime} \rangle
- \langle\varphi_{x^\prime}\varphi_{w^\prime} \rangle \langle\varphi_{y^\prime}\varphi_{z^\prime} \rangle
- \langle\varphi_{x^\prime}\varphi_{z^\prime} \rangle \langle\varphi_{y^\prime}\varphi_{w^\prime} \rangle \big)\,, \nonumber\\
&& =\frac{i}{2} G^{-1}_{xx^\prime}G^{-1}_{yy^\prime}G^{-1}_{ww^\prime}G^{-1}_{zz^\prime} \bigg(\langle \varphi_{x^\prime} \varphi_{y^\prime} \varphi_{w^\prime} \varphi_{z^\prime} \varphi_a \varphi_b\rangle
- G_{ab} \langle \varphi_{x^\prime} \varphi_{y^\prime} \varphi_{w^\prime} \varphi_{z^\prime}\rangle
-(6)G_{x^\prime y^\prime} \langle \varphi_a \varphi_b \varphi_{w^\prime} \varphi_{z^\prime}\rangle\ \nonumber\\
&&~~~~~~~~~~~~~~~~~~~+2(3) G_{a b} G_{x^\prime y^\prime} G_{wz}\bigg)\,,\nonumber\\
&& = \frac{1}{2}\bigg((8) G_{a z_1} \delta _{b z} V_{w x y z_1}+(4) G_{z_3 z_4} G_{a z_1} G_{b z_2} V_{x z_1 z_2 z_3} V_{w y z z_4}+ (6)   G_{z_3 z_4} G_{a z_1} G_{b z_2} V_{x z z_2 z_3} V_{w y z_1 z_4}\nonumber\\
&&~~~~~~+G_{a z_1} G_{b z_2} M_{z_1 z_2;w x y z}\bigg)\,.
   \eea
   Now we consider the contribution obtained when the derivative acts only on the inverse propagators. 
To differentiate the inverse propagators we use:
\bea
\label{chainX}
\frac{\delta G^{-1}_{ij}}{\delta R_{ab}} && = \frac{\delta G^{-1}_{ij}}{\delta G_{mn}}\frac{\delta G_{mn}}{\delta R_{ab}}\,,
\eea
with the result for $\delta G_{mn}/\delta R_{ab}$ given in equation (\ref{GderR2}) and the derivative of the inverse propagator given by:
\bea
\label{gert}
-2\frac{\delta G^{-1}_{ij}}{\delta G_{mn}} = G^{-1}_{im} G^{-1}_{jn} + G^{-1}_{in} G^{-1}_{jm}\,.
\eea  
Using equations (\ref{GderR2}), (\ref{chainX}) and (\ref{gert}) we obtain:
\bea
\label{VderR2part2}
V^c_{x^\prime y^\prime w^\prime z^\prime}\frac{\delta}{\delta R_{ab}}\bigg(G^{-1}_{xx^\prime}G^{-1}_{xx^\prime}G^{-1}_{xx^\prime}G^{-1}_{xx^\prime}\bigg) = -\frac{1}{2}\bigg((8) G_{a z_1} \delta _{b z} V_{w x y z_1}+(4) G_{z_3 z_4} G_{a z_1} G_{b z_2} V_{x z_1 z_2 z_3} V_{w y z z_4}  \bigg)\,. \nonumber\\ \,.
   \eea
Substituting the results in equations (\ref{VderR2part1}) and (\ref{VderR2part2}) into (\ref{VderR2-a}) we obtain:
\bea 
\label{VderR2}
\frac{\delta}{\delta R_{ab}} V_{xywz} = \frac{1}{2}G_{a z_1} G_{b z_2}\bigg(M_{z_1 z_2;w x y z} + (6) G_{z_3 z_4}  V_{x z z_2 z_3} V_{w y z_1 z_4}\bigg)\,.
\eea
The amputated derivative can be written:
\bea 
\label{VderR2-amp}
2\frac{\delta}{\overline{\delta} R_{ab}} V_{xywz} = \bigg(V_{w x y z ab} + (6) G_{z_3 z_4}  V_{x z z_2 z_3} V_{w y ab}\bigg) = \tilde M_{ab;xyzw}\,.
\eea

\end{document}